\documentclass[letterpaper, 10 pt, conference]{ieeeconf}  

\IEEEoverridecommandlockouts                              

\usepackage{graphicx}
\usepackage{url}  

\usepackage{amsmath}

\title{\LARGE \bf
A Low-Compexity Deep Learning Framework \\ For Acoustic Scene Classification
}

\author{Lam~Pham$^{1}$, 
        Hieu~Tang$^{2}$,
        Anahid~Jalali$^{1}$,
        Alexander~Schindler$^{1}$,
        Ross~King$^{1}$ \\ \\
        1. Center for Digital Safety \& Security, Austrian Institute of Technology, Austria. \\
2. Department of Electronic and Electrical Engineering, Hongik University, Korea. \\
        
}

\begin{document}

\maketitle
\thispagestyle{empty}
\pagestyle{empty}

\begin{abstract}

In this paper, we presents a low-complexity deep learning frameworks for acoustic scene classification (ASC). 
The proposed framework can be separated into three main steps: Front-end spectrogram extraction, back-end classification, and late fusion of predicted probabilities. 
First, we use Mel filter, Gammatone filter and Constant Q Transfrom (CQT) to transform raw audio signal into spectrograms, where both frequency and temporal features are presented.
Three spectrograms are then fed into three individual back-end convolutional neural networks (CNNs), classifying into ten urban scenes.
Finally, a late fusion of three predicted probabilities obtained from three CNNs is conducted to achieve the final classification result.
To reduce the complexity of our proposed CNN network, we apply two model compression techniques: model restriction and decomposed convolution. 
Our extensive experiments, which are conducted on DCASE 2021 (IEEE AASP Challenge on Detection and Classification of Acoustic Scenes and Events) Task 1A development dataset, achieve a low-complexity CNN based framework with 128 KB trainable parameters and the best classification accuracy of 66.7\%, improving DCASE baseline by 19.0\%.
\newline

\indent \textit{Key words}--- Convlutional neural network, Gammatone filter, constant Q transoform, MEL filter, spectrogram, deep learning.
\end{abstract}

\section{Introduction}
\label{intro}
\begin{figure*}[t]
	\centering
    \scalebox{0.8}{
	\centerline{\includegraphics[width=\linewidth]{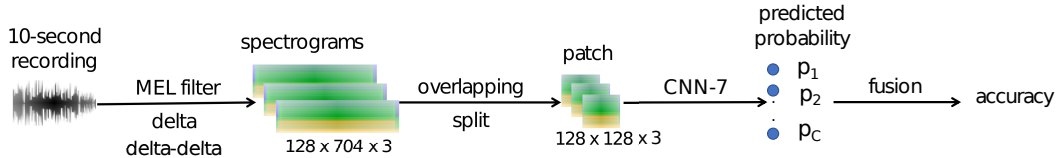}}
	}
	\vspace{0.1cm}
	\caption{High-level architecture of ASC baseline system proposed.}
	\label{fig:high_level_arc}
\end{figure*}
\begin{table*}[t]
    \caption{The CNN-7 network architecture baseline (input patch of $128{\times}128{\times}3$)} 
        	\vspace{-0.2cm}
    \centering
    \scalebox{1.0}{
    \begin{tabular}{|l |c|} 
        \hline 
            \textbf{Network architecture}   &  \textbf{Output}  \\
        \hline 
         BN - Convolution ([$3{\times}3] $@$  C_{out}1=32$) -  ReLU - BN - Dropout (10\%) & $128{\times}128{\times}32$\\
         BN - Convolution ([$3{\times}3] $@$  C_{out}2=32$) - ReLU - BN - AP [$2{\times}2$] - Dropout (10\%) & $64{\times}64{\times}32$\\
         
         BN - Convolution ([$3{\times}3] $@$ C_{out}3=64$) - ReLU - BN - Dropout (10\%) & $64{\times}64{\times}64$\\
         BN - Convolution ([$3{\times}3] $@$ C_{out}4=64$) - ReLU - BN - AP [$2{\times}2$] - Dropout (10\%) & $32{\times}32{\times}64$\\
         
         BN - Convolution ([$3{\times}3] $@$ C_{out}5=128$) - ReLU - BN - AP [$2{\times}2$] - Dropout (10\%) & $16{\times}16{\times}128$\\
         
         BN - Convolution ([$3{\times}3] $@$ C_{out}6=128$) - ReLU - BN - GAP - Dropout (10\%) & $128$\\
                  
         FC - Softmax  &  $C=10$       \\
       \hline 
    \end{tabular}
    }
    \label{table:VGG} 
\end{table*}

Acoustic Scene Classification (ASC), one of main research fields of `Machine Hearing'~\cite{lyon2017_bk}, has drawn much attention in recent years and has been applied to a wide range of real-life applications such as enhancing the listening experience of users by detecting scene context~\cite{ap1, ap2}, supporting sound event detection~\cite{ap3} or integrating in a multiple-sensor automatic system~\cite{ap4}.
To deal with ASC challenges such as unbalanced data, lacking data input, or mismatch recording devices, various methods have been proposed, which can be separated into two main approaches.
The first approach makes use of multiple data input such as ensemble of spectrograms~\cite{lam01, lam05, truc_dca_18} or audio channels~\cite{yuma}.
Meanwhile, the second approach focuses on back-end classification, and proposes powerful deep learning network architectures, which are able to enforce the training process~\cite{lam02, phaye_dca_18, zhao_dca_17, hong_dca_18}.
Although these two approaches can achieve good results, they present high-complexity systems.
Indeed, while multiple input data requires an ensemble of multiple individual classification models~\cite{lam03, lam04}, powerful network architectures show a number of convolutional layers~\cite{lam02, phaye_dca_18}.
All top-10 systems proposed in recent DCASE chalenges in  2018, 2019, 2020 are also based on complex architectures, requiring larger than 2 MB of trainable parameters.
This issue of network complexity prevents implementing edge devices concerning real-life applications which require a low footprint.
Although there are various methods proposed to deal with the issue of model complexity such as quantizaton~\cite{quant_01}, pruning~\cite{pruning_01, pruning_02}, model restriction (i.e. restriction on the number of layers~\cite{restrict_layer}, the number of kernel~\cite{restrict_width}, or both of these factors~\cite{restrict_all}), decomposed convolution~\cite{decompose_cnn}, hybrid methods using pruning and decomposed convolution~\cite{decompose_cnn}, pruning and distillation~\cite{hybid_comp_01}, these are mainly applied for image data. 
Therefore, our work introduces a low-complexity deep learning framework for ASC.
To deal with ASC challenges, we propose an ensemble of multiple spectrogram inputs, using Mel filter~\cite{librosa_tool}, Gammaton~\cite{aud_tool} filter, and CQT~\cite{librosa_tool}.
For each network used for training an individual spectrogram input, we deal with the issue of model complexity by combining model restriction and decomposed convolution methods.  

The remaining of our paper is organized as follows: Section 2 presents deep learning frameworks proposed and model compression techniques. Section 3 introduces evaluation setup where experimental setting, metric, and implementation of deep learning frameworks proposed are presented. Next, Section 4 presents and analyses the experimental results. Finally, Section 5 presents conclusion and future work.

\section{The low-complexity deep learning framework proposed}

\subsection{Our baseline}
\label{baseline}
We first propose a baseline with a high-level architecture shown in Fig.~\ref{fig:high_level_arc}, which comprises three main steps in the order of front-end spectrogram feature extraction, back-end classification, and a fusion of predicted probabilities.
In the first step, a raw audio signal is firstly transformed into a spectrogram of $129\times704$ by using MEL filter~\cite{librosa_tool} with Fast Fourier Transform (FFT) number, window size, and hope size set to 8192, 4096, and 620, respectively.
As delta and delta-delta across the temporal dimension are applied to the spectrogram, we then generate the spetrograms of $129\times704\times3$. 
Next, the spectrograms are split into 10 patches of $128\times128\times3$ with 50\% overlapping before feeding into a CNN based network for classification.

As we illustrate our proposed CNN based network architecture in Table~\ref{table:VGG}, it contains sub-blocks, which perform convolution with $C_{out}$ channel (Convolution ([kernel size]$@C_{out}$)), batch normalization (BN)~\cite{batchnorm}, rectified linear units (ReLU)~\cite{relu}, average pooling (AV [kernel size]), global average pooling (GAP), dropout (percentage dropped)~\cite{dropout}, fully-connected (FC), and Softmax layers. 
The dimension of Softmax layer is set to $C=10$ that equals to the number of scene context classified.
In total, we have 6 convolutional layers and 1 fully-connected layers that makes the proposed network architecture like CNN-7. 

As the CNN-7 works on patches, the final predicted probability of an entire spectrogram is computed by averaging of all patches.
Let us consider $\mathbf{P^{n}} = (\mathbf{p_{1}^{n}, p_{2}^{n},..., p_{C}^{n}})$,  with $C$ being the category number and the \(n^{th}\) out of \(N\) patches fed into the CNN-7, as the probability of a test instance, then the average classification probability is denoted as  \(\mathbf{\bar{p}} = (\bar{p}_{1}, \bar{p}_{2}, ..., \bar{p}_{C})\) where,
\begin{equation}
    \label{eq:mean_stratergy_patch}
    \bar{p}_{c} = \frac{1}{N}\sum_{n=1}^{N}p_{c}^{n}  ~~~  for  ~~ 1 \leq n \leq N 
\end{equation}
and the predicted label \(\hat{y}\) of the test instance evaluated is determined as:
\begin{equation}
    \label{eq:label_determine}
    \hat{y} = arg max (\bar{p}_{1}, \bar{p}_{2}, ...,\bar{p}_{C} )
\end{equation}

\subsection{Ensemble of multiple spectrogram inputs}
\label{late_fusion}
Although  both  of CQT and STFT spectrograms are built on Fourier Transform theory, they  extract  different  central  frequencies. 
Meanwhile, both Mel spectrogram and Gammatongram are generated from STFT spectrogram, but use two different filter banks: Mel and Gammatone filters.
We can conclude that three spectrograms either extract different central frequencies or apply different auditory models. 
Therefore, each spectrogram may contain its own distinct and complimentary information. 
This inspires us to propose an ensemble of these three spectrograms as a rule of thumb to improve the ASC performance~\cite{lam03, lam04}.
To evaluate the ensemble of multiple spectrograms, we proposed a late fusion scheme, referred to as PROD fusion.
In particular, we conduct experiments over individual networks with different spectrogram inputs, then obtain predicted probability of each network as  \(\mathbf{\bar{p_{s}}}= (\bar{p}_{s1}, \bar{p}_{s2}, ..., \bar{p}_{sC})\), where $C$ is the category number and the \(s^{th}\) out of \(S\) networks evaluated. 
Next, the predicted probability after PROD fusion \(\mathbf{p_{f-prod}} = (\bar{p}_{1}, \bar{p}_{2}, ..., \bar{p}_{C}) \) is obtained by:

\begin{equation}
\label{eq:mix_up_x1}
\bar{p_{c}} = \frac{1}{S} \prod_{s=1}^{S} \bar{p}_{sc} ~~~  for  ~~ 1 \leq s \leq S 
\end{equation}

Finally, the predicted label  \(\hat{y}\) is determined by (\ref{eq:label_determine}).

\subsection{Model compression methods applied to the CNN-7 network}

Our proposed CNN-7 architecture reports a complexity of 1,129 MB for non-zero parameters with using 32 bits for representing one trainable parameter. 
Additionally, using ensemble of three spectrogram inputs make the number of trainable parameters further increase three times.
To reduce the model complexity, we firstly restrict the number of channels used in the CNN-7 baseline, then reduce the channels of $C_{out}1$ from 32 to 16, $C_{out}3$ and $C_{out}4$ from 64 to 32, $C_{out}5$ and $C_{out}6$ from 128 to 64.
Our proposed channel restriction (CR) helps to reduce an individual CNN-7 complexity to 313 KB that nearly equals to 1/4 of the original size.

\begin{figure}[t]
	\centering
	\centerline{\includegraphics[width=\linewidth]{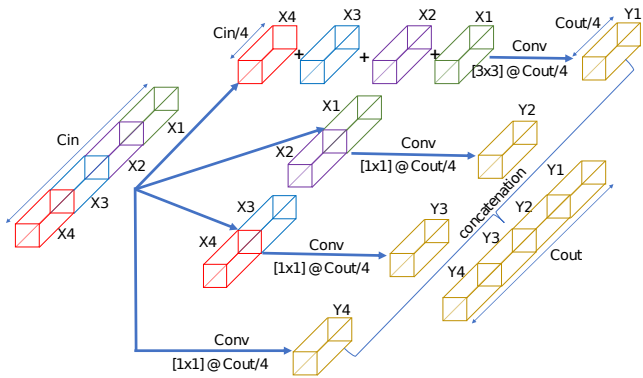}}
	\vspace{0.1cm}
	\caption{Decomposed convolution technique applied to a convolutional layer.}
	\label{fig:decompose}
\end{figure}

We further reduce the CNN-7 complexity by applying the decomposed convolution (DC) technique described in~\cite{decompose_cnn, dcase_decomp}.
Let us consider $C_{in}$ and $C_{cout}$ as the input and output channel numbers respectively, $W=3$ and $L=3$ are the dimensions of kernel size, which are used for a convolutional layer. 
Then, the number of trainable parameters at a convolutional layer is computed by $ W \times L \times C_{in} \times C_{out} = 9 \times C_{in} \times C_{out}$  .
We reduce the number of trainable parameters at a convolutional layer by decomposing the convolutional layer into 4 sub-convolutional layers as described in Fig.~\ref{fig:decompose}.
For all four sub-convolutional layers, the output channel is reduced to $C_{out}/4$.
Regarding the first sub-convolutional layer (the upper path shown in Fig. \ref{fig:decompose}), although we still use kernel size of [W$\times$L]=[3$\times$3], we reduce the input channels to $C_{in}$/4, then cost $(9 \times C_{in} \times C_{out})/16$ trainable parameters.
Regarding the other sub-convolution layers, we reduce the kernel size to [W$\times$L]=[1$\times$1]. 
While the second and third sub-convolutional layers (two middle paths shown in Fig. \ref{fig:decompose}), the input channel is reduced to $C_{in}/2$, it is remained in the fourth sub-convolutional layer (the lower path shown in Fig. \ref{fig:decompose}).
As a result, it requires $(C_{in} \times C_{out})/8$ for the second and third sub-convolutional layers, and  $(C_{in} \times C_{out})/4$ for the fourth sub-convolution layer.
By decomposing a convolutional layer into four sub-convolutional layers, the model complexity is reduced to nearly 1/8.5 of the original size.
By combining the two model compression techniques, we can achieve a CNN-7 network architecture with complexity of 42.6 KB, which nearly equals to 1/25 of the original size (i.e. the original CNN-7 network architecture is proposed in the baseline framework in Table~\ref{table:VGG}).
As we need to use three CNN-7 for three different spectrogram inputs, the final complexity of the proposed framework is approximately 128 KB.
%
\section{Evaluation Setting}

\subsection{TAU Urban Acoustic Scenes 2020 Mobile, development dataset~\cite{dc_2021_1A} (DCASE 2021 Task 1A)}
This dataset was proposed for DCASE 2021 challenge~\cite{dcase_web}, which requires a limitation of model complexity set to 128 KB with using 32 bits for one trainable parameter.
The dataset in slightly unbalanced, recorded from 12 large European cities: Amsterdam, Barcelona, Helsinki, Lisbon, London, Lyon, Madrid, Milan, Prague, Paris, Stockholm, and Vienna. 
It consists of 10 scene classes: airport, shopping mall (indoor), metro station (underground), pedestrian street, public square, street (traffic), traveling by tram, bus and metro (underground), and urban park.
The audio recordings were recorded from 3 different devices namely A (10215 recordings), B (749 recordings), C (748 recordings).
Additionally, synthetic data for 11 mobile devices was created based on the original recordings, referred to as S1 (750 recordings), S2 (750 recordings), S3 (750 recordings), S4 (750 recordings), S5 (750 recordings), and S6 (750 recordings). 

As a result, this task not only requires low complexity model, but it also proposes an issue of mismatch recording devices.  
To evaluate, we follow the DCASE 2021 Task 1A challenge~\cite{dcase_web}, separate this dataset into training (Train.) and evaluation (Eval.) subsets as shown in Table~\ref{table:dataset}.
Then, Train. subset is used for training frameworks proposed and Eval. subset is used for evaluating. 
Notably, two of 12 cities and S4, S5, S6 audio recordings are only presented in the Eval. subset for evaluating the issue of mismatch recording devices and unseen samples.

\subsection{Deep learning framework implementation}

We use Tensorflow framework to build all classification models in this papers.
The cross-entropy loss function used for training is described as

\begin{equation}
    \label{eq:loss_func}
    LOSS_{EN}(\Theta) = -\frac{1}{N}\sum_{n=1}^{N}\mathbf{y_n} \log \left\{\mathbf{\hat{y}_{n}}(\Theta) \right\} 
\end{equation}
defined over all parameters \(\Theta\), and $N$ is the number of training samples.  $\mathbf{y_{n}}$ and $\mathbf{\hat{y}_{n}}$  denote ground truth and predicted output.
The training is carried out for 100 epochs using Adam ~\cite{Adam} for optimization.

\subsection{Metric for evaluation}
\begin{table}[t]
	\caption{The number of 10-second audio-visual scene recordings corresponding to each scene categories in the Train. and Eval. subsets separated from the DCASE 2021 Task 1A development dataset~\cite{ds_2021_1b}.} 
	\vspace{-0.1cm}
	\centering
	\begin{tabular}{|l| c| c| } 
		\hline 
		\textbf{Category}         & \textbf{Train.}        & \textbf{Eval.}     \\ 
		\hline 
		Airport 	       & 1393                   & 296           \\        
		Bus     	       & 1400                   & 297           \\        
		Metro	 	       & 1382                   & 297           \\        
		Metro Station 	       & 1380                   & 297        \\        
		Park                   & 1429                   & 297        \\        
		Public square          & 1427                   & 297        \\        
		Shopping mall 	       & 1373                   & 297       \\        
		Street pedestrian      & 1386                   & 297          \\        
		Street traffic 	       & 1413                   & 297         \\        
		Tram 	               & 1379                   & 296          \\   \hline 
         Total              &  13962  & 2968  \\
                            &  ($\approx$38.79 hours) & ($\approx$8.25 hours) \\
		\hline 
	\end{tabular}    
	\label{table:dataset} 
\end{table}
Regarding the evaluation metric used in this paper, we follow DCASE 2021 challenge. 
Let us consider $C$ as the number of audio/visual test samples which are correctly classified, and the total number of audio/visual test samples is $T$, the classification accuracy (Acc. (\%)) is the \% ratio of $C$ to $T$.

\subsection{Optimize the framework proposed by evaluating factors of time length and data augmentation}
\begin{table*}[t]
    \caption{Performance comparison among three spectrograms with different time lengths, with or without using data augmentation} 
        	\vspace{-0.2cm}
    \centering
    \scalebox{1.0}{
    \begin{tabular}{| c |  c c c c  | c c c c |} 
        \hline 
        & \multicolumn{4}{c|}{\textbf{Without data augmentation}} & \multicolumn{4}{c|}{\textbf{Data augmentation}} \\ 
        \hline
            \textbf{Spectrogram}     &\textbf{1-second}  & \textbf{2-second}  &\textbf{5-second}  & \textbf{10-second} &  \textbf{1-second}  & \textbf{2-second} &  \textbf{5-second}  & \textbf{10-second}   \\

        \hline 
	    MEL     &56.7           &57.9           &56.2          &60.5           &54.6         &57.9           & 59.5          &58.4  \\
        GAM     &53.2           &55.0           &53.1          &53.9           &58.9         &60.1           & 60.6          &57.1 \\
        CQT     &44.3           &47.7           &48.6          &49.2           &44.2         & 45.7          & 48.6          &49.1  \\
        \hline                                                       
    \end{tabular}
    }
    \label{table:effect} 
\end{table*}
\begin{table}[t]
    \caption{Performance comparison among DCASE baseline, the CNN-7 baseline, the CNN-7 baseline with channel restriction (CNN-7 w/ CR), the CNN-7 baseline with channel restriction and decomposed convolution (CNN-7 w/ CR \& DC).} 
        	\vspace{-0.2cm}
    \centering
    \scalebox{1.0}{
    \begin{tabular}{| l |  c c c c  | } 
        \hline 
           &\textbf{DCASE}     &\textbf{CNN-7}    &\textbf{CNN-7}   &\textbf{CNN-7 w/}   \\
                           &\textbf{baseline}  &\textbf{baseline} &\textbf{w/ CR }  &\textbf{CR \& DC}   \\
         \textbf{Category}  &\textbf{(90.3 KB)}  & \textbf{(1.1 MB)}  &\textbf{(313 KB)}  & \textbf{(42.6 KB)}   \\
       \hline 

	    Airport           &40.5           &59.5           &50.3  &64.5          \\
        Bus               &47.1           &73.7           &70.4  &69.0          \\
        Metro             &51.9           &57.6           &49.8  &70.0         \\
        Metro station     &28.3           &53.9           &48.1  &45.1         \\
        Park              &69.0           &73.1           &78.5  &74.4         \\
        Public square     &25.3           &34.3           &38.4  &25.9         \\
        Shopping mall     &61.3           &52.9           &50.2  &43.4         \\
        Street pedestrian &38.7           &39.4           &35.0  &32.7         \\
        Street traffic    &62.0           &84.5           &88.2  &89.6         \\
        Tram              &53.0           &67.9           &62.5  &52.7         \\
        \hline                                                       
        Average           &47.7           &59.7           &57.1  &56.7         \\
        \hline                                                       

    \end{tabular}
    }
    \label{table:bs_cmp} 
\end{table}

In this paper, we further evaluate factors of time length and data augmentation which may affect the ASC performance, then find the most optimized framework.
In particular, we evaluate four different time lengths of 1 second (i.e. 1-second patch is used in the baseline proposed), 2 seconds, 5 seconds and 10 seconds (entire audio recording). 
In order to evaluate different time lengths mentioned but still keep the input patch of 128$\times$128$\times$3 unchanged, the hop size is set to 620, 1120, 1850 for 1-second, 2-second, and 5-second lengths respectively, while the other setting mentioned in Section~\ref{baseline} are unchanged.
To evaluate an entire 10-second recording, the hop size is set to 2048, then generate one patch of 128$\times$200$\times$3.

We enforce the back-end classifiers by applying two methods of data augmentation on the patches: mixup~\cite{mixup1, mixup2}, and spectrum agumentation~\cite{spec_aug}.
We then compare the ASC performance with and without using these data augmentation methods.
As we apply mixup data augmentation~\cite{mixup1, mixup2}, the labels of the mixup data input are no longer one-hot.
We therefore train back-end classifiers with Kullback-Leibler (KL) divergence loss~\cite{kl_loss} rather than the standard cross-entropy loss over all $N$ mixup training samples:
\begin{align}
    \label{eq:loss_func}
    LOSS_{KL}(\Theta) = \sum_{n=1}^{N}\mathbf{y}_{n}\log(\frac{\mathbf{y}_{n}}{\mathbf{\hat{y}}_{n}})  +  \frac{\lambda}{2}||\Theta||_{2}^{2},
\end{align}
where $\Theta$ denotes the trainable network parameters and $\lambda$ denote the $\ell_2$-norm regularization coefficient. $\mathbf{y_{c}}$ and $\mathbf{\hat{y}_{c}}$  denote the ground-truth and the network output, respectively. The training is carried out for 100 epochs using Adam~\cite{Adam} for optimization.

\section{Experimental Results And Discussion}
\subsection{Performance comparison between DCASE baseline and the CNN-7 baseline with or without using model compression methods}
As experimental results are shown in Table~\ref{table:bs_cmp}, although the CNN-7 baseline outperforms DCASE baseline and helps to improve the accuracy by 12\%, the CNN-7 baseline complexity is much larger than DCASE baseline.
By using model compression methods, we can achieve a low-complexity model referred to as CNN-7 with CR \& DC, which is nearly 1/2 of the DCASE baseline complexity, but still outperforms DCASE baseline showing an accuracy improvement of 9\%.

\subsection{Effect of time length, data augmentation, spectrogram input}
Next, we conduct experiments to evaluate effect of spectrogram input, time length, and data augmentation on the CNN-7 with CR \& DC, which are shown in Table~\ref{table:effect}.
As experimental results are shown in Table~\ref{table:effect}, MEL and GAM outperform CQT at different time lengths and with or without using data augmentation.
Without using data augmentation, MEL obtains better results than GAM at different time lengths. 
However, both MEL and GAM achieve competitive results with using data augmentation.
It can be seen that data augmentation is effective only for GAM.
Additionally, increasing time length helps to improve the accuracy for both CQT and MEL, but not much effective for GAM.
As a result, we finally configure an optimized and low-complexity framework for ASC task with setting:  CNN-7 with CR \& DC, 5-second time length, and using mixup \& spectrum data augmentation. 

\subsection{Evaluate ensemble of different spectrogram inputs}

\begin{figure*}[t]
    \centering
    \includegraphics[width =0.9\linewidth]{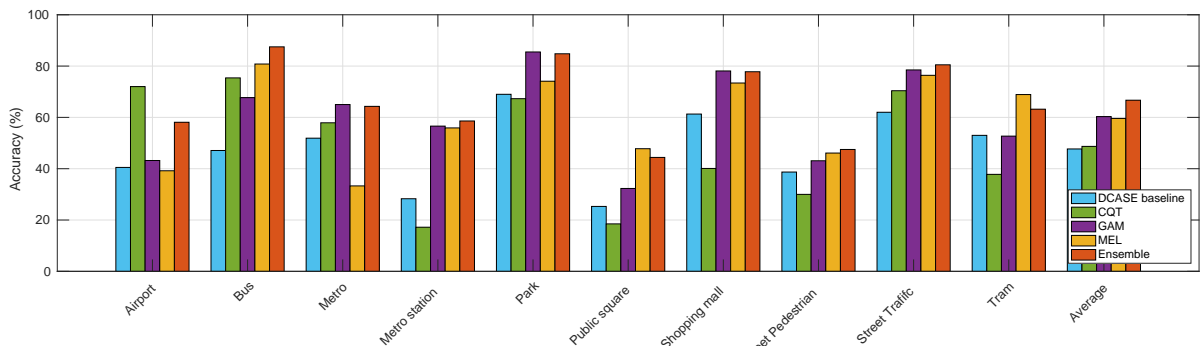}
    	\vspace{-0.5cm}
	\caption{Performance comparison (Acc.\%) of DCASE baseline, individual spectrograms (CQT, GAM, and MEL), and the ensemble of three spectrograms across all scene categories (using  CNN-7 with CR \& DC, 5-second time length, and mixup \& spectrum data augmentation for each spectrogram input)}
    \label{fig:F5}
\end{figure*}

Given the optimized framework, we conduct PROD fusion of three predicted probabilities from three spectrogram inputs (i.e. PROD fusion is mentioned in Section~\ref{late_fusion}) to obtain the final classification accuracy.
We then compare performances among DCASE baseline, the optimized framework with individual spectrograms, the optimized framework with the ensemble of multiple spectrograms, across all scene categories.
As experimental results are shown in Fig.\ref{fig:F5}, GAM and MEL achieve competitive results, and outperform CQT at almost scene categories except for 'Airport' and 'Bus'.
The ensemble of three spectrogram inputs helps to achieve an average accuracy of 66.7\%, improving DCASE baseline by 19.0\%, and notably showing improvement over all scene categories.

Further analysing performance over different recording devices as shown in Table~\ref{table:device}, we see that device A outperforms the other devices as this device is dominant in Train. subset. 
Although there is a lacking of training samples for device B and C, they achieves  competitive accuracy of 69.6\% and 70.8\% respectively, compared with device A performance of 79.1\%.
Regarding synthesized devices from S1 to S6, although there is no samples from S4, S5, S6 in Train. subset, the performance of these devices are competitive to the other S1, S2, S3.
Our results and analysis indicate that the ASC framework proposed not only shows low complexity of 128 KB, it also can tackle the issue of mismatched recording devices.

\section{Conclusion}

We have just presented a low-complexity framework for ASC, which makes use of multiple spectrogram inputs and model compression techniques.
While the ensemble of multiple spectrograms helps to tackle different ASC challenges of mismatch recording devices or lacking input to improve the ASC performance, a combination of model restriction and decomposed convolution techniques is effective to achieve a low model complexity of 128 KB. 
In the future, we will further compress the model complexity by combining our proposed approaches with other techniques of distillation, pruning, and quantization.

\begin{table}[t]
	\caption{The number of 10-second audio-visual scene recordings corresponding to each device in the Train. and Eval. subsets separated from the DCASE 2021 Task 1A development dataset~\cite{ds_2021_1b} and performance for each devices.} 
	\vspace{-0.1cm}
	\centering
	\begin{tabular}{|c| c| c| c| } 
		\hline 
		\textbf{Devices}         & \textbf{Train.}        & \textbf{Eval.}   & \textbf{Acc. \%}   \\ 
		\hline 
		A 	       & 10215                   & 330         &  79.1 \\        
		B     	       & 749                   & 329   &69.6         \\        
		C	 	       & 748                   & 329  &70.8          \\        
		S1 	       & 750                   & 330     &65.8    \\        
		S2                   & 750                   & 330   &63.6      \\        
		S3          & 750                   & 330    &67.0     \\        
		S4 	       & 0                   & 330  &63.9      \\        
		S5      &  0                  & 330       & 60.0   \\        
		S6 	       & 0                   & 330    & 60.3      \\        
		\hline 
	\end{tabular}    
	\label{table:device} 
\end{table}

\addtolength{\textheight}{-12cm}   

\bibliographystyle{IEEEbib}
\bibliography{refs}

\end{document}